\documentclass[nocite]{epl}
\usepackage{graphicx}
\usepackage{dcolumn}
\usepackage{bm}
\newcommand{\be}{\begin{equation}}
\newcommand{\ee}{\end{equation}}
\newcommand{\ba}{\begin{eqnarray}}
\newcommand{\ea}{\end{eqnarray}}

\begin{document}
\title{ On the dependence of the avalanche angle on the granular layer thickness}

\author{Matthieu Wyart}
\institute {School of Engineering and Applied Sciences, Harvard University, 29 Oxford Street, Cambridge, MA 02138}
\date{\today}

\begin{abstract}

A layer of sand of thickness $h$  flows down a rough surface if the inclination is larger than some threshold value $\theta_{start}$ which decreases with $h$.  A tentative microscopic model for the dependence of $\theta_{start}$ with $h$  is proposed for rigid frictional grains, based on the following hypothesis: (i) a horizontal layer of sand has some coordination $z$ larger than a critical value $z_c$ where mechanical stability is lost (ii) as the tilt angle is increased, the configurations visited present a growing proportion $z_{s}$ of sliding contacts.  Instability with respect to flow occurs when $z-z_s=z_c$. This criterion leads to a prediction for  $\theta_{start}(h)$ in good agreement with empirical observations. 
\end{abstract}

\maketitle

\begin{figure}
\vspace{-0.5cm}
\includegraphics[width=0.68\textwidth]{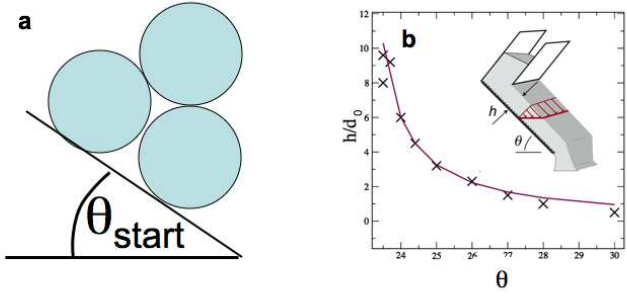}
\vspace{-1cm}
\caption{\label{fig1}  (a) A disc sitting on a regular array of similar discs is unstable if the tilt angle is larger than $\theta_{start}=30^0$. This simple picture cannot explain the dependence of $\theta_{start}$ with the thickness of the granular layer. (b)   Layer thickness in units of particle diameter  $h/d_0$  {\it v.s.} $\theta_{start}$ in an incline plane configuration where grains flow down a rough surface. Crosses correspond to the empirical results of Forterre and Pouliquen \cite{forterre}. The curve is the fit of the prediction Eq. (\ref{10}). }
\end{figure}

A layer of dry sand on an inclined plane cannot remain stable if it is sufficiently tilted: there exists an angle $\theta_{start}$ above which grains must flow. 
Considering a grain on the top of the layer, this observation makes perfect intuitive sense\cite{douady}: such a grain sits in a ``hole" made by the grains below,
as the system is tilted the center of gravity of this grain shifts in the direction of the tilt, and the grain must eventually escape its hole, as illustrated in Fig.(\ref{fig1}.a). 
Less intuitive is the observation \cite{pouliquen} that $\theta_{start}$ decreases with the thickness $h$ of the granular layer, and that this dependence can be observed for $h$ as large as ten particle diameters,
as illustrated in Fig.(\ref{fig1}.b). This implies that the stability of the layer is not governed by local rules: the grains at the top must be able to feel the presence of a fixed rough boundary significantly far below them. 
In general, one indeed expects that mechanical stability is a non-local property: imposing the stability of individual particles is a necessary condition, but is clearly not sufficient in general, as any
collective motion of the particles must also be stable. The non-local microscopic criterion on the stability of an assembly of particles is due to Maxwell \cite{max}. As we shall recall he showed that the key parameter 
is the coordination number $z$ (the average number of contacts per particle), which must satisfy some lower bound to allow stability. Here I shall extend this result to the situation where a fixed boundary (the rough surface of the inclined plane) and a free boundary (the upper layer of grains) are present. This result, together with the hypothesis supported by numerical analysis \cite{staron} that more and more contacts are sliding as the inclination increased,
leads to a reasonable estimate of $\theta_{start}(h)$.

I start by recalling Maxwell's criterion for the mechanical stability of $N$ frictionless particles in $d$ dimensions, forming $N_c$ contacts, whose coordination is $z\equiv 2N_c/N$.  
If $k_{ij}$ is the stiffness interaction, and if the  strain of the contacts is small \cite{matthieu2}, the energy expansion can be written as:
\be 
\label{1} 
\delta E=\sum_{\langle ij\rangle} \frac{k_{ij}}{2} [(\delta {\vec
R_i}-\delta {\vec R_j})\cdot {\vec n_{ij}}]^2 +o(\delta R^2)
\ee 
where  ${\vec n_{ij}}$ is the unit vector going from $i$ to $j$, and $\delta {\vec R_i}$ is the displacement of particles $i$. A system is mechanically stable if it
cannot be deformed without an energy cost, i.e. there exists no soft mode or particle displacement for which $\delta E=0$. Using Eq.(\ref{1}) this implies that the system of equation
\be 
\label{2}
(\delta {\vec R_i}-\delta {\vec R_j})\cdot {\vec n_{ij}}=0 \ \forall ij. 
\ee
must not present any non-trivial solution. Eq.(\ref{2}) is a linear system and non-trivial solutions must exist if it has more degrees of freedom than equations. There are $Nd$ degrees of freedom and $N_c$ equations, therefore stability requires $N_c> Nd$ or equivalently $z\geq 2d$, as derived by Maxwell. Thus, any solid made of frictionless particles must satisfy $z\geq2d$. It turns out that in the limit of large stiffness and fixed compression,
the system must become isostatic, i.e $z=2d$ \cite{moukarzel,tom1,roux}. This can be understood as follows: for infinite stiffness particles are just touching each other, therefore each pair of particles in contact must satisfy:
\be
\label{3}
||{\vec R_i}-{\vec R_j}||=d_0 \ \forall ij. 
\ee
where $d_0$ is the particle diameter. Again Eq.(\ref{3}) has $Nd$ degrees of freedom and $N_c$ constraint, it can therefore be satisfied only if $Nd\geq N_c$. Together with the stability condition $N_c\geq Nd$, this leads to $Nd=N_c$ or $z=2d$, as observed numerically \cite{J}. Thus frictionless rigid grains have a well-defined coordination.

For frictional particles the situation is different. As far as stability is concerned, there are now $d(d+1)/2$ degrees of freedom per particles ($d$ translations and $d(d-1)/2$ rotations), and each contact now brings $d$ constraints ( 1 due to the radial interaction and $d-1$ due to friction, corresponding to the directions transverse to the contact). The Maxwell criterion reads $z\geq d+1\equiv z_c$. In the stiff limit, the argument associated with Eq.(\ref{3}) is still valid, and therefore we get for the coordination:
\be
\label{4}
d+1\leq z\leq 2d
\ee
In frictional rigid grains the coordination is therefore not fixed, but depends on the system preparation, and on the friction coefficient $\mu$. Numerically \cite{roux2,somfai} it is found that as $\mu\rightarrow 0$ leads $z\rightarrow 2d$ (i.e when the friction coefficient vanishes one recovers the frictionless limit),
whereas $\mu \rightarrow \infty$ leads to $z\rightarrow d+1$. For example for poly-disperse discs with $\mu=0.5$  one finds numerically $z\approx 3.2$ \cite{somfai}.

In a frictional assembly of grains  some contacts can slip: the ratio of transverse to longitudinal force in such contacts is $\mu$. These  contacts  display no more transverse stiffness, they do not bring transverse restoring force when an external load is applied. If $N_s$ is the number of slipping contacts, the system displays $N_s (d-1)$ less constraints than if no contacts were slipping. We shall assume that slipping contacts leave  the number of degrees of freedom relevant for the mechanical stability unchanged (in the discussion section we shall argue that this assumption is valid for intermediate and large friction coefficient, but not for nearly frictionless particles). Introducing:
\be
\label{5a}
 z_s\equiv 2(d-1) N_s/N,
 \ee
 the Maxwell counting arguments now leads to:
\be
\label{5}
z\geq z_c+z_s
\ee
Under increasing shear, or when a granular layer is tilted, one expects for the configurations visited that the ratio of the typical transverse contact force to normal force will grow. This leads to an increase in the number of contacts that slip. In the limit of rigid grains the dependence of $z_s$ on the applied stress can only be a function of the ratio $\sigma/p=tan(\theta)$, where $p$ is the pressure and $\sigma$ the shear stress:
\be
\label{6}
z_s=f_\mu(tan(\theta))
\ee
Numerical computation of $f_\mu(tan(\theta))$ in two dimensions for $\mu=0.5$ are available\cite{staron}.
Eq.(\ref{5}) and Eq.(\ref{6}) leads to the following condition on $\theta$ for the stability of an infinitely thick layer of grains ($h=\infty$):
\be
\label{7a}
f_\mu(tan(\theta))\leq z-z_c
\ee
 The coordination $z$ may a priori depends on $\theta$. We shall assume that this is not the case, as supported by numerics \cite{staron}. Relaxing this hypothesis would lead to no qualitative changes in the following. 
We also assume that a layer of sand starts flowing when the bound(\ref{7a}) is saturated:
\be
\label{7}
f_\mu(tan (\theta_{start}(h=\infty)))\approx z-z_c
\ee
This assumes that the destabilization induced by slipping contacts occurs at an angle where other destabilizing effects (such as the local instability described in Fig.(\ref{fig1}.a)) did not yet set in. 
As far as orders of magnitude are concerned our  assumption is supported by the numerical studies  \cite{staron,somfai}  observing respectively $f_\mu(tan(\theta_{start}))\approx 0.16$ and $z-z_c\approx 0.2$ for polydisperse discs of friction $\mu=0.5$. In other words, Eq.(\ref{7}) leads to a reasonable estimate of $\theta_{start}$ in those numerics.

\begin{figure}
\vspace{-0.5cm}
\includegraphics[width=0.68\textwidth]{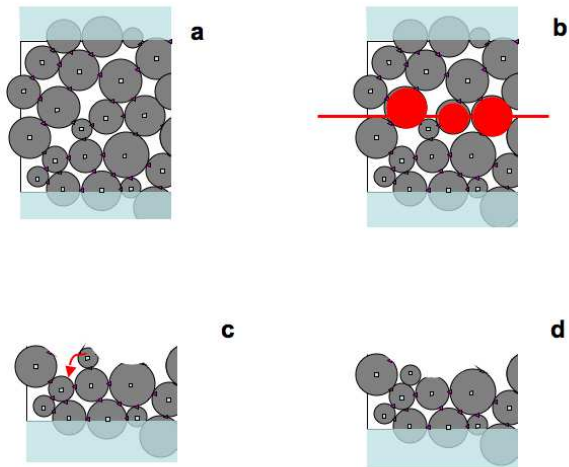}
\vspace{-1cm}
\caption{\label{fig2}  (a) Assembly of frictionless particles with periodic boundary conditions (b)  Model of a rough substrate: all the particles intersecting an arbitrary horizontal line, represented in red, are pinned. This increases the stability of the system, in a manner equivalent to an increase of coordination, as computed in the text. (c) Naive model  of a free interface, generated by removing all the red particles  crossing the arbitrary line. This model is not realistic because some of the particle are left with less than $d$ contacts, and are therefore unstable. The presence of such an interface decreases the stability of the system, and is equivalent to a decrease of coordination. The effect is exactly opposite to the presence of a fixed boundary. (d) A more realistic free interface is generated by letting the unstable particles at the surface flow and make $d$ contacts. This increases the coordination. As a consequence the destabilizing effect of a free boundary is milder than what expected from the naive model represented in (c), and the overall effect of the joined presence of a free and fixed boundary is stabilizing.}
\end{figure}

We know extend this result to the case of finite thickness, and first show that the presence of a fixed boundary (such as the rough plane on which grain sits) increases the coordination by some amount $a d_0/h$, where $h$ is the thickness of system and $a$ is a constant of order one. For concreteness consider a system of linear size $h$ of $N$ frictionless particles with periodic boundary conditions (the argument also holds with friction for any geometry), such as the one shown in Fig.\ref{fig2}.a. If $m$ disjoined particles are pinned, the system formed by the other particles loses $md$ degrees of freedom, but still present the same number of contacts.  The ratio of constraints per particle has therefore increased: this is equivalent to  increasing the coordination by an amount $\Delta z=2[N_c/(N-m) -N_c/N] \approx 2(N_c/N) (m/N)=z m/N$, where we have assumed $m/N<<1$. To mimic the presence of a rough surface we can pin all the particles crossing a surface of the system, such as the red line represented in Fig.\ref{fig2}.b.  In this case a fraction of order $m\sim d_0/h$ of the particles are pinned, and the coordination increases by some amount $\Delta z = a d_0/h$ with respect to the coordination $z$ of a bulk solid.

To study the effect of a free boundary we shall instead consider the situation where $m$ particles are removed of the system.  The remaining system  looses again $md$ degrees of freedom, but it also looses $mz$ contacts. As far as degrees of freedom and constraints are considered, this is equivalent to decrease the coordination by some amount $\Delta z=2[(N_c-mz)/(N-m)-N_c/N]\approx-am/N$. To mimic the presence of a free boundary, we can again draw a surface and remove all the particles crossing it, this procedure creates a free surface, as shown in Fig.\ref{fig2}.c. From this simple argument we may (too) naively expect that the effect of a free boundary is exactly the opposite a fixed boundary: it decreases the  coordination by $\Delta z= -a d_0/h$. If so, the finite size effect of a  layer of grains with a free and a fixed boundary would simply cancel out. This is nevertheless not the case, because as it appears in Fig.\ref{fig2}.c,  removing a layer of particles does not lead to a realistic description of a free interface: in particular a finite fraction of the particles are left with less than $d$ contacts, an unrealistic situation. A better description of a free interface is obtained by letting such particles flow down, until they make $d$ contacts, as exemplified in Fig.\ref{fig2}.d.  Since this occurs for a finite fraction of the particles at the interface, the coordination of the system increases by some amount of order $d_0/h$. Consequently the presence of a free boundary decreases the effective coordination by an amount $\Delta z=a' d_0/h$, with $a'<a$.

Accounting for these effects, Eq.(\ref{7}) can be generalized to describe a layer of grains with finite thickness:
\be
\label{8}
f_\mu(tan (\theta_{start}))=z-z_c+(a-a')d_0/h
\ee
It is convenient to expand $f_\mu(tan(\theta))$ in the vicinity of $\theta_{start}(h=\infty)$ where the value of $f_\mu$ is known from Eq.(\ref{7}):
\be
\label{9}
f_\mu(tan (\theta))\approx (z-z_c) + a_\mu (tan(\theta)-tan(\theta_{start}(\infty)))
\ee
where $a_\mu$ is the derivative of $f_\mu$ at $\theta_{start}(\infty)$. $a_\mu$ is expected  to be a decreasing function of $\mu$, as a larger friction coefficient will decrease the rate at which the number
 of slipping contacts increases. 
Using Eq.(\ref{9}), Eq.(\ref{8}) can be rewritten as:
\be
\label{10}
\frac{h}{d_0}=\frac{a-a'}{a_\mu(tan( \theta_{start}(h))-tan (\theta_{start}(\infty)))}
\ee
which is our final result. Once the value of $\theta_{start}(\infty)$ is fixed (and as discussed above, Eq.(\ref{7}) leads to a reasonable estimate of this quantity in the numerical systems where it can be tested),
Eq.(\ref{10}) has one remaining fitting parameter $(a-a')/a_\mu$, which we expect to be an increasing function of the friction coefficient. Fig.\ref{fig1}.b shows that the form proposed in Eq.(\ref{10}) fits well the empirical observations. 

I conclude with several remarks:

(i) The model proposed here may apply to a rough plane, such as the one obtained by gluing particles on a surface. If the interaction grain-plane is  instead described by a friction coefficient $\mu_s$ significantly smaller than $\mu$,
one expects that the first contacts to slide will be at the interface grain-surface, and that the granular layer will slip as a bloc for some angle determined by $\mu_s$, and $\theta_{start}$ should not depend significantly on $h$. At the other extreme where the plane is extremely rough, the grains in contact with the plane may be significantly immobilized. If so the present result can still be applied replacing $h$ by $h-d_0$. 

(ii) Our analysis neglects spatial heterogeneities in the coordination and in the location where contacts slip, and is in some sense a mean-field model. If large fluctuations in these quantities occur, some regions may yield for angles smaller than our estimate.
Small avalanches are indeed observed as the inclination increases, and become more frequent close to $\theta_{start}$ \cite{kabla}. Our analysis will hold qualitatively if such avalanches cannot trigger a global flow except near the critical inclination for which the system as a whole becomes unstable.

(iii) In a shear cell where the two boundaries are fixed, the same line of thought leads to an equation similar to Eq.(\ref{10}), where $a-a'$ is replaced by $2a$. One thus expects the finite size effects to be stronger in this geometry.

(iv)  When deriving Eq.(\ref{5}), we have assumed that the contacts that slip leave the number of degrees of freedom relevant for mechanical stability unchanged. This is true except when the fraction $\nu=f_\mu/z$ of such contacts is large. In that case, all the contacts of a particle may slip, with a probability of order $\nu^z$. Rotating  such a particle leaves the contact forces unchanged. Thus as far as  linear response is concerned these particles act as frictionless particles, and their degrees of freedom of rotation is irrelevant for mechanical stability. This implies that Eq.(\ref{5}) over-estimates the destabilizing effects of the slipping contacts. Nevertheless for intermediate and large friction coefficient, the corresponding corrections are negligible. For example in numerics \cite{staron} with $\mu=0.5$ it was observed that $\nu\approx 0.05$ and $z\approx 3$, and we expect the number of particles acting as frictionless ones to be of order $0.05^3= 1.25*10^{-4}<<1$. Nevertheless, this effect must be dominant if frictionless particles, such as emulsions, are considered. Then  one expects the system to remain isostatic ($z=2d$) independently of $\theta$, leading to $f_\mu=0$. For emulsions the nature of the instability at  $\theta_{start}$ must therefore differ from the present description.

(iv) Once the flow is initiated, it stops when the slope of the granular surface becomes smaller than some angle $\theta_{stop}<\theta_{start}$. Empirically $\theta_{stop}$ depends on $h$ in a way qualitatively similar to $\theta_{start}$ \cite{pouliquen}, and it seems unlikely that these two similar dependences stem from very different causes. This suggests that the properties of dense granular flows are affected by the presence of nearly stable configurations, an idea with long history in the context of the glass transition \cite{goldstein}.

It is a pleasure to thank B. Andreotti, A. Kabla and O. Pouliquen for discussions and comments on the manuscript.

\end{document}